\documentclass[12pt]{article}
\usepackage{amsmath,epsfig,color}

\begin{document}
\pagestyle{empty}


\vskip 2cm
\begin{center}
{\Large\bf Generalized index theorem  \\
for topological superconductors with
Yang-Mills-Higgs couplings
}
\end{center}

\vspace*{1cm}
\vspace*{1cm}
\def\thefootnote{\fnsymbol{footnote}}
\begin{center}{\sc Takanori Fujiwara and 
Takahiro Fukui}
\end{center}
\vspace*{0.2cm}
\begin{center}
{\it Department of Physics, Ibaraki University,
Mito 310-8512, Japan}
\end{center}

\vfill
\begin{center}
{\large\sc Abstract}
\end{center}
\noindent

We investigate an index theorem for a Bogoliubov-de Gennes 
Hamiltonian (BdGH) describing a topological superconductor
with Yang-Mills-Higgs couplings in arbitrary dimensions. 
We find that the index of the BdGH is determined solely by 
the asymptotic behavior of the Higgs fields and is 
independent of the gauge fields.
It can be nonvanishing  if the dimensionality of the order parameter 
space is equal to the spatial dimensions. In the presence 
of point defects there appear localized zero energy states
at the defects. Consistency of the index with the existence 
of zero energy bound states is examined explicitly in a vortex 
background in two dimensions and in a monopole 
background in three dimensions. 
\vfil

\newpage
\pagestyle{plain}

\section{Introduction}
\label{sec:intro}
\setcounter{equation}{0}

Yang-Mills-Higgs systems admit topologically 
nontrivial field configurations such as vortices 
and magnetic monopoles \cite{nielsen, thooft}. They are considered to be 
responsible for various nonperturbative effects 
in particle physics. Recently topological 
insulators and superconductors have attracted 
much interest as a new phase of materials in 
condensed matter physics. Topological insulators
are band insulators whose ground states are characterized by topological numbers \cite{TKNN}. 
The method of topologically classifying 
insulating ground states has been generalized to
superconducting states described by the mean-field 
Bogoliubov-de Gennes Hamiltonian (BdGH). This provides a unified framework
of the classification of non-interacting fermion systems
with respect to 
time reversal and particle-hole 
symmetries \cite{Zirnbauer,AZ,SRFL,Kitaev}. The 
topological classification was further extended to 
systems with topological defects \cite{TK}, 
which enables us to classify zero modes along line defects and zero energy bound states
localized at point defects from the point of view of the topological universality class.

For a superconductor with time-reversal symmetry
(class BDI), 
there exists a unitary operator that anticommutes with the BdGH. 
It defines an extended chiral symmetry. 
Generically, chiral symmetry ensures the topological stability 
of the zero energy states, since
these states are controlled by the index 
theorem \cite{Callias, Weinberg,NiemiSemenoff} for the BdGH, implying 
the robustness 
under continuous deformations of the order parameters. 
Remarkably, the extended chiral symmetry can 
be defined in arbitrary dimensions.
This is a sharp contrast to the usual chiral invariance in 
massless Dirac theories, 
which only exists in even dimensions. The extended 
chiral invariance also incorporates internal spin 
symmetry. By gauging the internal spin symmetry 
we can reformulate the system as a Yang-Mills-Higgs 
system, where the space varying gap parameters can 
be regarded as Higgs fields. 

In this paper we investigate the index theorem of the Dirac-type
BdGH corresponding to the Yang-Mills-Higgs systems in 
arbitrary dimensions with special emphasis on the extended 
chiral invariance. 
Previously, zero energy bound states in two and three dimensions
were investigated by Jackiw and Rebbi \cite{Jackiw-Rebbi}, and
Jackiw and Rossi \cite{Jackiw-Rossi}. The index theoretical 
approaches of their models were explored by Callias \cite{Callias} 
and Weinberg \cite{Weinberg}.
In particular, 
the Weinberg index theorem turned out to be quite useful for much more 
complicated systems such as non-Abelian vortices in a color superconductor \cite{ffny}. 
We show that for more generic models with Yang-Mills-Higgs couplings in arbitrary dimensions
the Weinberg index theorem is applicable, and
the index of the BdGH can be 
nonvanishing, implying that there appear topological zero energy bound 
states localized at the point 
defect if the dimensions of order parameter space is 
equal to the spatial dimensions. We also see 
that the index is solely determined by the behaviors of 
the Higgs fields at the spatial infinities. This is 
expected in odd dimensions \cite{Callias}, since we have no topological 
invariant defined by the gauge field. In even dimensions, 
however, we have two topological invariants, one defined 
by the Higgs fields and one by the Yang-Mills fields. 
The expression of the topological invariants 
seems to contain both of them. One might consider that the number of 
zero modes would be affected by including the 
Yang-Mills fields. We show that this is not the case. 
These two topological invariants are so combined that a 
unique index is obtained. This was noted for vortex 
background in two dimensions \cite{Weinberg}. We show 
this happens in arbitrary dimensions.

This paper is organized as follows. In Sect. \ref{sec:bdgh} we 
introduce BdGH in arbitrary dimensions 
and their generalized chiral symmetry. We then define 
the generalized index for the BdGH. A 
computation of the topological index is given 
in Sect. \ref{sec:ccd}. The chiral current at the spatial 
infinities is investigated in Sect. \ref{sec:cindH}. General 
properties of the chiral current and the relation
with the topological index in even dimensions are 
given in Sect. \ref{sec:dga}. Zero modes for vortex and 
monopole configurations in two and three dimensions 
are investigated in Sect. \ref{sec:tttd}. Finally, 
Sect. \ref{sec:sd}. is devoted to summary and discussions.

\section{Extended chiral symmetry of 
BdGH}
\label{sec:bdgh}
\setcounter{equation}{0}

We begin with a $d$ dimensional fermion system coupled 
to an O($d$) Higgs field $\phi_a$ $(a=1,\cdots, d)$. 
We assume that the system is described by the following 
Hamiltonian 
\begin{eqnarray}
  \label{eq:h}
  H_0&=&-i\gamma^j\partial_j+\phi(x), \qquad 
  (\phi(x)=\Gamma^a\phi_a(x))
\end{eqnarray}
where $\gamma^j$ ($j=1,\cdots,d$) and $\gamma^{d+a}
=\Gamma^a$ form a set of $2d$ dimensional $\gamma$ 
matrices satisfying $\{\gamma^\mu,\gamma^\nu\}=\delta^{\mu\nu}$. 
We introduce the $2d$ dimensional chiral matrix 
by $\gamma_{2d+1}=(-i)^d\gamma^1\cdots\gamma^{2d}
=(-i)^d\gamma^1\cdots\gamma^d\Gamma^1\cdots\Gamma^d$. 
It anti-commutes with both $\gamma^j$ and $\Gamma^a$. The 
system then possesses the chiral symmetry in the sense 
that 
\begin{eqnarray}
  \label{eq:csymh}
  \{\gamma_{2d+1},H_0\}&=&0. 
\end{eqnarray}
Note that the chiral symmetry can be defined in any 
dimensions. This is contrasted with the usual chiral 
symmetry which is only defined in even dimensions. 
The Hamiltonian (\ref{eq:h}) concerns itself with $2d$ 
dimensions, $d$ spatial and $d$ internal.
The chiral invariance is known to be violated 
in the presence of the chemical potential. In the 
present work we are concerned with the chiral 
symmetric case, where the index of the Hamiltonian 
is well-defined. 

In addition to the chiral invariance we can 
define particle-hole symmetry. Let us introduce the 
charge conjugation matrix $C$ by
\begin{eqnarray}
  \label{eq:C}
  C(\gamma^j)^\ast C^{-1}=\gamma^{j}, \qquad
  C(\Gamma^a)^\ast C^{-1}=-\Gamma^a. 
\end{eqnarray}
It is always possible to find $C$ for a given set of 
$\gamma^\mu$. Then $H_0$ satisfies 
\begin{eqnarray}
  \label{eq:phs}
  {\cal C}H_0{\cal C}^{-1}&=&-H_0, 
\end{eqnarray}
where ${\cal C}=CK$ denotes complex conjugation $K$  
followed by the multiplication by $C$. To each 
particle state $\psi$ of an energy $E>0$ we can 
define a state $\psi_c$ with energy $-E$ by 
\begin{eqnarray}
  \label{eq:phconj}
  \psi_c(x)={\cal C}\psi(x)=C\psi^\ast(x). 
\end{eqnarray}
Eq. (\ref{eq:phs}) ensures the particle-hole 
symmetry of the spectrum of $H_0$. 

Before going into detailed analysis of the index 
of the Hamiltonian we introduce spin($d$) gauge field 
\begin{eqnarray}
  \label{eq:gf}
  A_j(x)&=&\frac{1}{2}\Sigma^{ab}A_{abj}(x), 
\end{eqnarray}
where $\Sigma^{ab}\equiv[\Gamma^a,\Gamma^b]/4$ 
are the spin($d$) generators and $A_{abj}$ are 
the components of the gauge field satisfying 
$A_{abj}^\ast=A_{abj}=-A_{baj}$. 
The gauge symmetric generalization of $H_0$ 
is then given by 
\begin{eqnarray}
  \label{eq:gsh}
  H&=&-i\gamma^jD_j+\phi(x),
\end{eqnarray}
where $D_j=\partial_j+A_j$ is the covariant 
derivative. For the Higgs field it is given
\begin{eqnarray}
  \label{eq:cdHig}
  D_j\phi=\Gamma^aD_j\phi_a, \quad
  D_j\phi_a=\partial_j\phi_a+A_{abj}\phi_b. 
\end{eqnarray}
The field strength is also defined as usual
\begin{eqnarray}
  \label{eq:Fabj}
  F_{ij}&=&\frac{1}{2}\Sigma^{ab}F_{abij}, \quad
  F_{abij}=\partial_iA_{abj}-\partial_jA_{abi}
  +A_{aci}A_{cbj}-A_{acj}A_{cbi}. 
\end{eqnarray}
Since the spin($d$) generators commute 
with $\gamma_{2d+1}$ and ${\cal C}$, the chiral 
and particle-hole symmetries remain intact by 
the generalization. 

Unlike Nielsen-Olsen vortex and 't Hooft-Polyakov 
monopole, which will be treated in Sect. \ref{sec:tttd}, 
we consider the Yang-Mills and Higgs fields as 
independent classical background. 
For the present we only assume that the gauge potential approaches 
to pure gauge and the Higgs field satisfies $\phi^2
\rightarrow|\phi_0|^2$ at the spatial infinities, 
where $|\phi_0|^2$ is a nonvanishing constant. 
This gives rise to a finite energy gap.
Due to the particle-hole symmetry mentioned above, 
any eigenstate of nonvanishing energy $E$ is 
necessarily paired 
with an eigenstate of energy $-E$. These states 
are also related with the chiral transformation. 
The zero modes of $H$, however, can be unpaired. 
They can be made self-conjugate under the particle-hole 
symmetry and can be regarded as Majorana 
states. For a general Yang-Mills-Higgs background 
it is not possible to find 
an explicit form of zero mode wave function. 
In the chirally symmetric case, however, we can 
investigate the existence or nonexistence of 
zero modes by computing the index of $H$ defined 
by 
\begin{eqnarray}
  \label{eq:indH}
  \mathrm{index}H&=&n_+-n_-,
\end{eqnarray}
where $n_\pm$ are the numbers of positive 
and negative $\gamma_{2d+1}$ chirality zero energy states. The 
index of $H$ is known to be a topological 
invariant, {\it i.e.}, it is invariant under continuous 
deformations of the gauge and Higgs fields. The 
index theorem relates the index with a topological 
invariant of these fields \cite{Callias,Weinberg,NiemiSemenoff}. 

To establish the index theorem we rewrite 
Eq.(\ref{eq:indH}) as
\begin{eqnarray}
  \label{eq:index}
  {\rm index}H
  &=&\lim_{m\rightarrow0}{\rm Tr}\gamma_{2d+1}\frac{m^2}{H^2+m^2} 
  \nonumber \\
  &=&\lim_{m\rightarrow0}\int d^dx\lim_{y\rightarrow x}
  {\rm tr}\gamma_{2d+1}\frac{m^2}{H^2+m^2}\delta^d(x-y), 
\end{eqnarray}
where $\mathrm{Tr}$ stands for the integration over 
the spatial coordinates as well as the trace on 
the $\gamma$ matrices. As is well-known, index 
of the Dirac operator is related with the chiral anomaly 
of the axial current divergence. To show this we introduce 
an axial current $J^i(x)$ by 
\begin{eqnarray}
  \label{eq:axc}
  J^i(x)&=&\lim_{m\rightarrow0\atop M\rightarrow\infty}\lim_{y\rightarrow x}
  {\rm tr}\gamma_{2d+1}\gamma^i\left(\frac{1}{iH+m}
    -\frac{1}{iH+M}\right)\delta^d(x-y), 
\end{eqnarray}
where $M$ is a Pauli-Villars mass to regularize the 
current at short distances. It is straightforward to 
show that the axial current divergence can be cast into 
the form 
\begin{eqnarray}
  \label{eq:acd}
  \partial_iJ^i(x)&=&-2\lim_{m\rightarrow0\atop M\rightarrow\infty}
  \lim_{y\rightarrow x}{\rm tr}\gamma_{2d+1}
  \left(\frac{m^2}{H^2+m^2}
    -\frac{M^2}{H^2+M^2}\right)\delta^d(x-y).
\end{eqnarray}
At this stage we can take the two limits, $m\rightarrow0$ and 
$M\rightarrow\infty$, separately. Eq.(\ref{eq:index}) 
then can be written as 
\begin{eqnarray}
  \label{eq:id}
  \mathrm{index}H&=&-\frac{1}{2}\int_{S_\infty^{d-1}} dS_iJ^i(x)+c_d, 
\end{eqnarray}
where $c_d$ is the topological index defined by 
\begin{eqnarray}
  \label{eq:cd}
  c_d&=&\lim_{M\rightarrow\infty}\mathrm{Tr}\gamma_{2d+1}
  \frac{M^2}{H^2+M^2}   
\end{eqnarray}
and $S_\infty^{d-1}$ denotes the infinities of the $d$ 
dimensional euclidean space.
In the case of index theorem on compact manifolds without 
boundaries the contribution from the chiral current on 
the rhs of (\ref{eq:id}) vanishes and the index coincides 
with $c_d$. Since we are working with the Hamiltonian (\ref{eq:gsh}) 
defined on a euclidean space, the surface term can be 
nontrivial \cite{Callias, Weinberg,NiemiSemenoff}. We expect a nonvanishing 
contribution to $\mathrm{index}H$ if $J^i(x)$ is of order 
$|x|^{-d+1}$ for $|x|\rightarrow\infty$. To compute the 
integral on the rhs of Eq. (\ref{eq:id}) we only need the 
leading behavior of the chiral current at spatial infinities. 
This will be done in Sect. \ref{sec:cindH}.

\section{Topological index}
\label{sec:ccd}
\setcounter{equation}{0}

We have shown that the index 
can be written as a sum of the topological index (\ref{eq:cd}) 
and the surface integral of the chiral current. The evaluation 
of the functional trace on the rhs of (\ref{eq:cd}) is similar 
to that of chiral anomalies. 
In this section we compute $c_d$. 

Eq. (\ref{eq:cd}) can be 
explicitly written as  
\begin{eqnarray}
  \label{eq:compcd}
  c_d&=&\lim_{M\rightarrow\infty}\int d^dx\lim_{y\rightarrow x}
  {\rm tr}\gamma_{2d+1}\frac{M^2}{H^2+M^2}\delta^d(x-y) \nonumber\\
  &=&\lim_{M\rightarrow\infty}\int d^dx
  \int \frac{d^dk}{(2\pi)^d}
  e^{-ikx}{\rm tr}\gamma_{2d+1}\frac{M^2}{H^2+M^2}
  e^{ikx}.
\end{eqnarray}
Using 
\begin{eqnarray}
  \label{eq:ekH}
  e^{-ikx}He^{ikx}
  &=&-i\gamma^i(D_j+ik_j)+\phi, \\
  \label{eq:ekHH}
  e^{-ikx}H^2e^{ikx}
  &=&-(D_j+ik_j)^2-\frac{1}{2}\gamma^i\gamma^jF_{ij}
  -i\gamma^jD_j\phi+\phi^2, 
\end{eqnarray}
we can compute the rhs of Eq. (\ref{eq:compcd}) as 
\begin{eqnarray}
  \label{eq:cdcomp}
  c_d&=&\lim_{M\rightarrow\infty}\int d^dx
  \int \frac{d^dk}{(2\pi)^d}{\rm tr}\gamma_{2d+1}
  \frac{M^2}{-(D_j+ik_j)^2-\frac{1}{2}\gamma^i\gamma^jF_{ij}
    -i\gamma^jD_j\phi+\phi^2+M^2} \nonumber \\
  &=&\lim_{M\rightarrow\infty}M^2\int d^dx
  \int \frac{d^dk}{(2\pi)^d}\sum_{n=0}^\infty
  \mathrm{tr}\Biggl[\gamma_{2d+1}
  \frac{1}{-(D_i+ik_i)^2+\phi^2+M^2} \nonumber \\
  &&\times
  \left(\left(i\gamma^iD_i\phi+\frac{1}{2}\gamma^i\gamma^jF_{ij}\right)
    \frac{1}{-(D_i+ik_i)^2+\phi^2+M^2}\right)^n\Biggr].
\end{eqnarray}
Due to the presence of $\gamma_{2d+1}$ the terms 
with $n<d/2$ vanish under the 
trace, whereas the terms with $n>d/2$ do not contribute 
to the sum in the limit $M^2\rightarrow\infty$ as can be easily 
seen by the scaling argument of the momentum variables 
$k\rightarrow Mk$. This immediately gives 
$c_d=0$ in odd dimensions. 

For $d=2N$ even we obtain 
\begin{eqnarray}
  \label{eq:cdfn}
  c_d&=&\int d^dx\int \frac{d^{2N}k}{(2\pi)^d}
  \frac{1}{(k^2+1)^{N+1}}
  \mathrm{tr}\gamma_{2d+1}
  \left(\frac{1}{2}\gamma^i\gamma^jF_{ij}\right)^N \nonumber \\
  &=&\frac{(-1)^N}{(2\pi)^NN!}\int d^dx
  \epsilon^{i_1\cdots i_d}\mathrm{tr}_\eta
  F_{i_1i_2}\cdots F_{i_{d-1}i_d}, 
\end{eqnarray}
where $\epsilon^{i_1\cdots i_d}$ is the Levi-Civita 
symbol in $d$ dimensions. We have also 
introduced $\mathrm{tr}_\eta$ by 
\begin{eqnarray}
  \label{eq:trh}
  \mathrm{tr}_\eta(\cdots)&=& \mathrm{tr}(\eta\cdots),
\end{eqnarray}
where $\eta$ is defined by 
\begin{eqnarray}
  \label{eq:eta}
  \eta&=&\frac{(-1)^{d(d-1)/2}}{2^d}\Gamma^1\cdots\Gamma^d.  
\end{eqnarray}
The overall normalization is chosen 
to give 
\begin{eqnarray}
  \mathrm{tr}_\eta\Gamma^{a_1}\cdots \Gamma^{a_d}=\epsilon^{a_1\cdots a_d}.
\end{eqnarray}
It satisfies anti-cyclic property 
\begin{eqnarray}
    \mathrm{tr}_\eta \Gamma\Gamma'=-\mathrm{tr}_\eta\Gamma'\Gamma 
  \quad \hbox{for}\quad \{\eta,\Gamma\}=0.
\end{eqnarray}
Later we need to compute traces involving $\phi$. In even dimensions $\eta$ anti-commutes with $\phi$, whereas it commutes in odd dimensions. 

We have seen that only the gauge field contributes to the topological 
index and the Higgs field is irrelevant to the computation of $c_d$. 
The result (\ref{eq:cdfn}) coincides with computation of index of 
a Dirac operator without Higgs fields.

\section{Computation of chiral current}
\label{sec:cindH}
\setcounter{equation}{0}

The chiral current defined by (\ref{eq:axc}) is written by 
the regularized fermion propagator. It is in general 
a nonlocal quantity of the background fields. To find its contribution to the 
index of the BdGH we only need the asymptotic behaviors 
at spatial infinities, where the chiral current turns 
out to become local. In this section we evaluate the 
asymptotic form of the chiral current. 

We first rewrite the current (\ref{eq:axc}) as 
\begin{eqnarray}
  J^i(x)&=&-i\lim_{m\rightarrow0\atop M\rightarrow\infty}
  \int\frac{d^dk}{(2\pi)^d}e^{-ikx}
  {\rm tr}\gamma_{2d+1}\gamma^iH\left(\frac{1}{H^2+m^2}
    -\frac{1}{H^2+M^2}\right)e^{ikx}.
\end{eqnarray}
The computation of $J^i(x)$ is similar to the one presented 
in Sect. \ref{sec:ccd}. We obtain 
\begin{eqnarray}
  \label{eq:Jiexp}
  J^i(x)&=&-i\lim_{m\rightarrow0\atop M\rightarrow\infty}
  \int\frac{d^dk}{(2\pi)^d}\sum_{n=0}^\infty
  \Biggl\{\frac{1}{(k^2+|\phi_0|^2+m^2)^{n+1}}
  -(m^2\rightarrow M^2)\Biggr\} \nonumber\\
  &&\times\mathrm{tr}\gamma_{2d+1}
  \gamma^i(\gamma^jk_j-i\gamma^jD_j+\phi)
  \left(i\gamma^kD_k\phi+\frac{1}{2}\gamma^k\gamma^lF_{kl}
    +\Delta\right)^n,
\end{eqnarray}
where $\Delta$ is given by
\begin{eqnarray}
  \label{eq:D}
  \Delta&=&2ik_iD_i+D_i^2-|\phi|^2+|\phi_0|^2.
\end{eqnarray}

At the spatial infinities 
$\phi^2$ approaches to a constant $|\phi_0|^2$. We 
further assume 
\begin{eqnarray}
  \label{eq:asympb}
  \phi^2-|\phi_0|^2, ~\partial_j\phi, ~ A_j 
  &\sim&\mathrm{O}(|x|^{-1})\quad\hbox{for}\quad 
  |x|\rightarrow\infty .
\end{eqnarray}
To find $\mathrm{index}H$ it is only 
necessary to know the leading  
$\mathrm{O}(|x|^{-d+1})$ terms of the current for 
$|x|\rightarrow\infty$. It is easy to convince oneself that the terms 
of the rhs of Eq. (\ref{eq:Jiexp}) vanish 
for $n<(d-1)/2$ because of the trace with 
$\gamma_{2n+1}$, whereas the terms with $n\geq d$ 
can be ignored since they decay faster than 
$|x|^{-d+1}$ for $|x|\rightarrow\infty$. 
Eq. (\ref{eq:Jiexp}) then can be written as 
\begin{eqnarray}
  J^i(x)&=&-i\lim_{m\rightarrow0\atop M\rightarrow\infty}
  \int\frac{d^dk}{(2\pi)^d}\sum_{n\geq (d-1)/2}^{d-1}
  \Biggl\{\frac{1}{(k^2+|\phi_0|^2+m^2)^{n+1}}
  -(m^2\rightarrow M^2)\Biggr\} \nonumber\\
  &&\times\mathrm{tr}\gamma_{2d+1}
  \gamma^i\phi
  \left(i\gamma^iD_i\phi+\frac{1}{2}\gamma^i\gamma^jF_{ij}\right)^n
  +\mathrm{O}(|x|^{-d}).
\end{eqnarray}
The $k$ integral can be done for $n\geq (d-1)/2$ as
\begin{eqnarray}
  \int\frac{d^dk}{(2\pi)^d}\frac{1}{(k^2+\mu^2)^{n+1}}
  &=&\frac{\Gamma(n+1-d/2)}{(4\pi)^{d/2}n!}\frac{1}{\mu^{2(n+1)-d}}. 
\end{eqnarray}
Now the limits $m\rightarrow0$ and $M\rightarrow\infty$ can be 
taken safely. Keeping only the nonvanishing contributions, 
we obtain
\begin{eqnarray}
  \label{eq:Jisym}
  J^i(x)
  &=&-i\sum_{l=0}^{\left[\frac{d-1}{2}\right]}
  \frac{\Gamma(d/2-l)}{(4\pi)^{d/2}(d-l-1)!}
  \mathrm{tr}\gamma_{2d+1}
  \gamma^i\hat\phi
  \mathrm{Symm}\left[\left(i\gamma^jD_j\hat\phi\right)^{d-2l-1}
    \left(\frac{1}{2}\gamma^j\gamma^kF_{jk}\right)^l\right]
  +\mathrm{O}(|x|^{-d}), \nonumber \\
\end{eqnarray}
where $\hat\phi=\phi/|\phi_0|$ and $\mathrm{Symm}$ denotes 
symmetrized product defined by
\begin{eqnarray}
  \mathrm{Symm}A^nB^m
  =\frac{(n+m)!}{n!m!}\left.\frac{\partial^{n+m}}{\partial s^n\partial t^m}
  \exp[sA+tB]\right|_{s=t=0}.
\end{eqnarray}
Since $\gamma^jD_j\hat\phi$ and $\gamma^j\gamma^kF_{jk}$ 
are effectively commutative in the trace of Eq. (\ref{eq:Jisym}), 
we can simplify the symmetrized product further. 
We thus arrive at the asymptotic form of the 
chiral current 
\begin{eqnarray}
  \label{eq:ji}
  J^i(x)&=&\frac{(-1)^{d-1}}{(4\pi)^{d/2}}
    \sum_{k=0}^{\left[\frac{d-1}{2}\right]}
    \frac{2^{d-k}\Gamma(d/2-k)}{k!(d-2k-1)!}\epsilon^{ii_2\cdots i_d}
    \mathrm{tr}_\eta\hat\phi D_{i_2}\hat\phi\cdots
    D_{i_{d-2k}}\hat\phi F_{i_{d-2k+1}i_{d-2k+2}}
    \cdots F_{i_{d-1}i_d}, \nonumber  \\
\end{eqnarray}
where nonleading contributions are suppressed. 
The asymptotic chiral current is local 
with respect to the background fields.
Unlike the topological index $c_d$ it exists in odd as well as even dimensions. 

It is easy to obtain explicit forms of $J^i(x)$ in low 
dimensions.  
In two dimensions the chiral current and topological 
index are given by 
\begin{eqnarray}
  \label{eq:Jitd}
  J^i(x)&=&\frac{1}{\pi}\epsilon^{ij}\epsilon^{ab}
  \hat\phi_a\partial_j\hat\phi_b
  -\frac{1}{2\pi}\epsilon^{ij}\epsilon^{ab}A_{abj}, \\
  \label{eq:cdtd}
  c_2&=&-\frac{1}{4\pi}\int d^2x\epsilon^{ij}
  \epsilon^{ab}\partial_iA_{abj}. 
\end{eqnarray}
These lead to the index 
\begin{eqnarray}
  \label{eq:indtd}
  \mathrm{index}H&=&\frac{1}{2\pi}\int_{S_\infty^1}dS_i
  \epsilon^{ij}\epsilon^{ab}\hat\phi_a\partial_j\hat\phi_b. 
\end{eqnarray}
We see that the topological index is canceled by 
the gauge field dependent term of the chiral current \cite{Weinberg}. 

Similar thing also happens in three dimensions.  
The topological index vanish identically in odd 
dimensions and the 
gauge field dependent terms of the chiral current 
can be cast into a total derivative term as 
\begin{eqnarray}
  \label{eq:Jithd}
  J^i(x)&=&-\frac{1}{4\pi}\epsilon^{ijk}\epsilon^{abc}
  \left(\hat\phi_aD_j\hat\phi_bD_k\hat\phi_c
    +\frac{1}{2}\hat\phi_aF_{jkbc}\right) \nonumber\\
  &=&-\frac{1}{4\pi}\epsilon^{ijk}\epsilon^{abc}\hat\phi_a\partial_j
  \hat\phi_b\partial_k\hat\phi_c
  -\frac{1}{4\pi}\epsilon^{ijk}\epsilon^{abc}
  \partial_j(A_{abk}\hat\phi_c).
\end{eqnarray}
We thus obtain the index 
\begin{eqnarray}
  \label{eq:indthd}
  \mathrm{index}H&=&\frac{1}{8\pi}\int_{S_\infty^2} dS_i
  \epsilon^{ijk}\epsilon^{abc}\hat\phi_a\partial_j
  \hat\phi_b\partial_k\hat\phi_c.
\end{eqnarray}
Again the gauge field dependence disappears 
and the index is only determined by the 
asymptotic behavior of the Higgs field 
at the infinities \cite{Callias}. In the next section we show 
that this holds true in arbitrary dimensions. 

\section{Differential geometric approach}
\label{sec:dga}
\setcounter{equation}{0}

To establish the gauge field independence 
of $\mathrm{index}H$ it is convenient to introduce 
Lie algebra valued differential forms \cite{zwz}, 
the gauge potential 1-form and the field strength 
2-form, as
\begin{eqnarray}
  \label{eq:gp}
  &&A=\frac{1}{4}\Gamma^a\Gamma^b\mathrm{d}x^iA_{abi}, \\
  \label{eq:fs}
  &&F=\frac{1}{2}\mathrm{d}x^i\mathrm{d}x^j
  F_{ij}
  =\frac{1}{8}\mathrm{d}x^i\mathrm{d}x^j\Gamma^a\Gamma^b
  F_{abij}
  =\mathrm{d}A+A^2. 
\end{eqnarray}
Note that $F$ satisfies Bianchi identity
\begin{eqnarray}
  \label{eq:bi}
  DF=dF+AF-FA=0. 
\end{eqnarray}
The exterior covariant 
derivative of $\hat\phi$ is given by 
\begin{eqnarray}
  \label{eq:Dp}
  D\hat\phi=\mathrm{d}\hat\phi+[A,\hat\phi]. 
\end{eqnarray}
Taking $D$ once more we obtain
\begin{eqnarray}
  \label{eq:DDp}
  D^2\hat\phi&=&[F,\hat\phi]. 
\end{eqnarray}
Since the generators of spin($d$) commutes 
with $\eta$ defined by (\ref{eq:eta}), so 
do $A$ and $F$ in arbitrary dimensions. 

The topological index (\ref{eq:cdfn})
in $d=2N$ dimensions can be written in terms 
of $F$ as 
\begin{eqnarray}
  \label{eq:cddf}
  c_d&=&\frac{(-1)^N}{\pi^NN!}\int \mathrm{tr}_\eta F^N,  
\end{eqnarray}
where the integral is taken over the entire 
$d$ dimensional space. If we introduce Chern-Simons 
form $\omega_{d-1}^0$ by
\begin{eqnarray}
  \label{eq:CSf}
  \mathrm{tr}_\eta F^N&=&\mathrm{d}\omega_{d-1}^0, 
\end{eqnarray}
the topological index (\ref{eq:cddf}) can be 
converted to the surface integral 
\begin{eqnarray}
  \label{eq:cdwn}
  c_d&=&\frac{(-1)^N}{\pi^NN!}\int_{S_\infty^{d-1}}\omega^0_{d-1},
\end{eqnarray}
where $S_\infty^{d-1}$ denotes the $(d-1)$-sphere at the spatial 
infinities.
The Chern-Simons form $\omega^0_{d-1}$ can be written as  
\begin{eqnarray}
  \label{eq:efcsf}
  \omega^0_{d-1}&=&N\int_0^1dt \mathrm{tr}_\eta A
  (t\mathrm{d}A+t^2A^2)^{N-1}.
\end{eqnarray}
For $d=2,4,6$ the Chern-Simons forms are explicitly given by 
\begin{eqnarray}
  &&\omega^0_1=\mathrm{tr}_\eta A, \nonumber \\
  &&\omega^0_3=\mathrm{tr}_\eta\left(A\mathrm{d}A+\frac{2}{3}A^3\right),
  \nonumber \\
  &&\omega^0_5=\mathrm{tr}_\eta\left(A(\mathrm{d}A)^2
    +\frac{3}{2}A^3\mathrm{d}A+\frac{3}{5}A^5\right). 
\end{eqnarray}

We now turn to the chiral current (\ref{eq:ji}). 
The first integral in the rhs of Eq. (\ref{eq:id}) can be 
written as 
\begin{eqnarray}
  \label{eq:dfsinf}
  \int_{S_\infty^{d-1}}dS_iJ^i&=&\int_{S_\infty^{d-1}}
  {}^\ast J,
\end{eqnarray}
where ${}^\ast J$ is the Hodge dual of the chiral current 
1-form $J=J_i(x)\mathrm{d}x^i$ and is given by
\begin{eqnarray}
  \label{eq:astJ}
  {}^\ast J&=&\frac{1}{(d-1)!}\epsilon_{i_1i_2\cdots i_d}J^{i_1}(x)
  \mathrm{d}x^{i_2}\cdots
  \mathrm{d}x^{i_d}. 
\end{eqnarray}
For the current Eq. (\ref{eq:ji}), the dual can 
be written as  
\begin{eqnarray}
  \label{eq:astdf}
  {}^\ast J&=&\sum_{k=0}^{\left[\frac{d-1}{2}\right]}
  C_{d,k}\mathrm{tr}_\eta\hat\phi(D\hat\phi)^{d-2k-1}F^k,
\end{eqnarray}
where $C_{d,k}$ is defined by 
\begin{eqnarray}
  \label{eq:Cdk}
  C_{d,k}&=&-(-1)^{\left[\frac{d+1}{2}\right]}
  \frac{\Gamma(d/2-k)}{\pi^{d/2}k!(d-2k-1)!}. 
\end{eqnarray}
Taking the exterior derivative of the current and
using Eq. (\ref{eq:bi}) and (\ref{eq:DDp}), 
we get 
\begin{eqnarray}
  \label{eq:dJa}
  \mathrm{d}{}^\ast J&=&\sum_{k=0}^{\left[\frac{d-1}{2}\right]}
  C_{d,k}\mathrm{tr}_\eta((D\hat\phi)^{d-2k}
  +\hat\phi[F,\hat\phi](D\hat\phi)^{d-2k-2}
  -\hat\phi D\hat\phi[F,\hat\phi](D\hat\phi)^{d-2k-3} \nonumber\\
  &&
  +\cdots+(-1)^{d-2k-2}\hat\phi(D\hat\phi)^{d-2k-2}
  [F,\hat\phi])F^k \nonumber\\
  &=&\sum_{k=0}^{\left[\frac{d-1}{2}\right]}
  C_{d,k}\mathrm{tr}_\eta(D\hat\phi)^{d-2k}F^k
  -\sum_{k=0}^{\left[\frac{d-1}{2}\right]}
  (d-2k-1)C_{d,k}\mathrm{tr}_\eta
  [F,\hat\phi]\hat\phi(D\hat\phi)^{d-2k-2}F^k. \nonumber \\
\end{eqnarray}
In deriving this use has been made of the fact 
that $[F,\hat\phi]$ effectively anti-commutes 
with $\hat\phi$ and $D\hat\phi$ in the trace. 
We can simplify the trace in the second summand 
on the rhs of Eq. (\ref{eq:dJa}) as 
\begin{eqnarray}
  \mathrm{tr}_\eta
  [F,\hat\phi]\hat\phi(D\hat\phi)^{d-2k-2}F^k&=&
  -\frac{1}{k+1}\mathrm{tr}_\eta\hat\phi(D\hat\phi)^{d-2k-2}([F,\hat\phi]F^k
  +F[F,\hat\phi]F^{k-1}\cdots+F^k[F,\hat\phi]) \nonumber \\
  &=&-\frac{1}{k+1}\mathrm{tr}_\eta
  \hat\phi(D\hat\phi)^{d-2k-2}[F^{k+1},\hat\phi] \nonumber \\
  &=&(-1)^d\frac{2}{k+1}\mathrm{tr}_\eta(D\hat\phi)^{d-2k-2}F^{k+1}.
\end{eqnarray}
Inserting this into Eq. (\ref{eq:dJa}), we obtain 
\begin{eqnarray}
  \label{eq:dJaf}
  \mathrm{d}{}^\ast J
  &=&-\left.\frac{2(d-2k+1)}{k}C_{d,k-1}\mathrm{tr}_\eta
    (D\hat\phi)^{d-2k}F^k\right|_{k=\left[\frac{d-1}{2}\right]+1},
\end{eqnarray}
where use has been made of 
$\mathrm{tr}_\eta(D\hat\phi)^d=0$ and the relation 
\begin{eqnarray}
  C_{d,k}=\frac{2(d-2k+1)}{k}C_{d,k-1} , \qquad
  \left(k=1,2,\cdots,\left[\frac{d-1}{2}\right]\right) .
\end{eqnarray}
The former can be verified by noting $\hat\phi^2=1$. 

In odd dimensions the coefficient on the rhs of 
(\ref{eq:dJaf}) vanishes. We therefore obtain 
\begin{eqnarray}
  \mathrm{d}{}^\ast J&=&0. 
\end{eqnarray}
This implies that the current can be written as
\begin{eqnarray}
  \label{eq:odaJ}
  -\frac{1}{2}{}^\ast J
  &=&\frac{(-1)^\frac{d+1}{2}}{2^d\pi^{\frac{d-1}{2}}(\frac{d-1}{2})!}
  \mathrm{tr}_\eta\hat\phi(\mathrm{d}\hat\phi)^{d-1}
  +\mathrm{d}\Omega_{d-2}(\hat\phi,A), 
\end{eqnarray}
where $\Omega_{d-2}$ is a $(d-2)$-form. 

In even ($d=2N$) dimensions Eq. (\ref{eq:dJaf}) 
can be written as 
\begin{eqnarray}
  \mathrm{d}{}^\ast J&=&2\frac{(-1)^N}{\pi^NN!}\mathrm{tr}_\eta F^N.
\end{eqnarray}
This is reminiscent of chiral anomaly. 
It also implies that ${}^\ast J$ can be written as
\begin{eqnarray}
  \label{eq:edaJ}
  -\frac{1}{2}{}^\ast J&=&(-1)^\frac{d}{2}
  \frac{\left(\frac{d}{2}\right)!}{\pi^\frac{d}{2}d!}
  \mathrm{tr}_\eta\hat\phi(\mathrm{d}\hat\phi)^{d-1}
  +\mathrm{d}\Omega_{d-2}(\hat\phi,A)
  -\frac{(-1)^\frac{d}{2}}{\pi^\frac{d}{2}\left(\frac{d}{2}\right)!}
  \omega_{d-1}^0. 
\end{eqnarray}
Interestingly enough, there appears the the Chern-Simons 
form. 

It is straightforward to check Eq. (\ref{eq:odaJ}) 
and (\ref{eq:edaJ}) in low dimensions. We give 
explicit expressions for $d=2,~3,~4,~5$:
\begin{eqnarray}
  d=2: &&{}^\ast J=\frac{1}{\pi}\mathrm{tr}_\eta\hat\phi\mathrm{d}\hat\phi
  -\frac{1}{\pi}\mathrm{tr}_\eta A, \nonumber \\
  d=3: &&{}^\ast J=-\frac{1}{4\pi}
  \mathrm{tr}_\eta\hat\phi(\mathrm{d}\hat\phi)^2
  -\frac{1}{\pi}\mathrm{d}\mathrm{tr}_\eta\hat\phi A, \nonumber\\
  d=4: &&{}^\ast J=-\frac{1}{6\pi^2}\mathrm{tr}_\eta
  \hat\phi(\mathrm{d}\hat\phi)^3
  +\frac{1}{\pi^2}\mathrm{d}\mathrm{tr}_\eta
  \left(\hat\phi\mathrm{d}\hat\phi A
    +\frac{1}{2}\hat\phi A\hat\phi A\right)
  +\frac{1}{\pi^2}\mathrm{tr}_\eta\left(A\mathrm{d}A
    +\frac{2}{3}A^3\right), \nonumber \\
  d=5: && \nonumber \\ 
  &&\hskip -2cm {}^\ast J=
  \frac{1}{32\pi^2}\mathrm{tr}_\eta\hat\phi(\mathrm{d}\hat\phi)^4
  +\frac{1}{4\pi^2}\mathrm{d}\mathrm{tr}_\eta\Biggl\{
  \hat\phi(\mathrm{d}\hat\phi)^2A
  +\hat\phi\mathrm{d}\hat\phi A\hat\phi A
  +\frac{1}{3}(\hat\phi A)^3
  +\hat\phi(A\mathrm{d}A+\mathrm{d}AA+A^3)\Biggr\}. \nonumber \\
\end{eqnarray}

We now turn to $\mathrm{index}H$. We see from 
Eq. (\ref{eq:odaJ}) or (\ref{eq:edaJ}) that $\Omega_{d-2}$ 
does not contribute to the index and the topological 
index in even dimensions is canceled by the Chern-Simons 
form. We thus obtain 
\begin{eqnarray}
  \label{eq:fit}
  \mathrm{index}H&=&(-1)^{\left[\frac{d+1}{2}\right]}
  \frac{\Gamma(d/2+1)}{\pi^{d/2}d!}\int_{S_\infty^{d-1}} \mathrm{tr}_\eta
  \hat\phi(\mathrm{d}\hat\phi)^{d-1}. 
\end{eqnarray}
The gauge fields disappear completely and the index is determined 
only by the behaviors of the Higgs fields at the spatial infinities. This 
generalizes the observation of Ref. \cite{Weinberg} for the two dimensional 
model of Jackiw and Rossi \cite{Jackiw-Rossi}. In the next section 
we will introduce Yukawa coupling constant $g$ by substituting 
$\phi$ by $g\phi$ in Eq. (\ref{eq:gsh}). In even dimensions the 
index (\ref{eq:fit}) is not affected by this, whereas an extra 
overall factor $\mathrm{sgn}(g)$ appears on the rhs of  Eq. 
(\ref{eq:fit}) in odd dimensions.

\section{Topological configurations in two and three dimensions}
\label{sec:tttd}
\setcounter{equation}{0}

So far we have considered the Higgs and gauge 
fields as independent background fields. In some 
model field theories topological objects can be realized 
dynamically as a solution to the field equations. More 
specifically we assume the gauge-Higgs system in $d$ spatial 
dimensions with a static energy 
\begin{eqnarray}
  \label{eq:sgHe}
  {\cal H}&=&\int d^dx\left(\frac{1}{8e^2}(F_{abij})^2
    +\frac{1}{2}(D_i\phi_a)^2
    +\frac{\lambda_0}{8}(|\phi_0|^2-|\phi|^2)^2\right),
\end{eqnarray}
where $e$ is the gauge coupling constant. 
It becomes stationary for the fields satisfying 
\begin{eqnarray}
  \label{eq:sfeq}
  D_iD_i\hat\phi_a+\lambda(1-|\hat\phi|^2)\hat\phi_a=0, 
  \qquad  D_iF_{abij}+\kappa\hat\phi_a\overleftrightarrow D_j\hat\phi_b=0, 
\end{eqnarray}
where $\kappa=e^2|\phi_0|^2$ and $\lambda=\lambda_0|\phi_0|^2/2$. 
These nonlinear field equations give rise to topological 
objects, Nielsen-Olesen vortex in two dimensions \cite{nielsen} 
and 't Hooft-Polyakov monopole  in three 
dimensions \cite{thooft}. We consider the BdG equation with these 
topological configurations and see more closely how the 
index relation is fulfilled. 

\subsection{Vortex in Maxwell-Higgs system}
\label{sec:vmhs}

In two dimensions Eq. (\ref{eq:sfeq}) can describe 
vortices. Ansatz for a vortex with unit vorticity is 
given by 
\begin{eqnarray}
  \label{eq:vans}
  \hat\phi_a(x)=h(r)\frac{x^a}{r}, \qquad
  A_{abi}(x)=-\epsilon_{ab}\epsilon_{ij}(1-k(r))\frac{x^j}{r^2},
\end{eqnarray}
where $h(r)$ and $k(r)$ are assumed to satisfy the 
following boundary conditions
\begin{eqnarray}
  \label{eq:BChkv}
  h(0)=0, \quad k(0)=1, \quad h(\infty)=1, \quad
  k(\infty)=0. 
\end{eqnarray}
Eqs. (\ref{eq:sfeq}) give differential equations 
for $h(r)$ and $k(r)$ as
\begin{eqnarray}
  \label{eq:hkeqv}
  h''+\frac{h'}{r}=\frac{k^2h}{r^2}-\lambda(1-h^2)h, \qquad
  k''-\frac{k'}{r}=\kappa h^2k. 
\end{eqnarray}
We see that $1-h$ and $k$ decrease exponentially for large $r$. 

We now turn to $\mathrm{index}H$ for the vortex background. 
The covariant derivative $D_i\hat\phi_a$ is given by 
\begin{eqnarray}
  \label{eq:Dfv}
  D_i\hat\phi_a&=&h'\frac{x^ix^a}{r^2}
  +\epsilon_{ij}\epsilon_{ab}hk\frac{x^jx^b}{r^3}.
\end{eqnarray}
It decays exponentially at $r\rightarrow\infty$, so does the 
chiral current $\ast J$. We see that $\mathrm{index}H$ 
coincides with the topological index $c_2$ since the chiral 
current has no contribution to the index. It is 
easy to compute $c_2$. The field strength can be found as 
\begin{eqnarray}
  \label{eq:Fv}
  F_{abij}&=&-\epsilon_{ab}\epsilon_{ij}\frac{k'}{r} .
\end{eqnarray}
Eq. (\ref{eq:cdwn}) then immediately gives 
$\mathrm{index}H=c_2=-1$.

This might be felt contradictive with the result of 
Sect. \ref{sec:dga} that $\mathrm{index}H$ is determined 
completely by the asymptotic behavior of the Higgs field. 
It is of course not the case. The gauge field is related with the 
Higgs field by the field equations. In particular the 
chiral current vanishes exponentially as $r\rightarrow\infty$. 
Therefore, the contribution from the Higgs current, 
the first term on the rhs of (\ref{eq:Jitd}), cancels 
that from the gauge current, the second term of the same 
equation, which in turn exactly cancels the topological 
index of the gauge field strength (\ref{eq:Fv}). This 
implies that the topological invariant (\ref{eq:indtd}) 
coincides with the topological index. 

The nonvanishing index suggests that the Hamiltonian given 
by Eq. (\ref{eq:gsh}) has one negative chirality zero mode. 
For the vortex 
background we can find zero mode  
of $H$ explicitly. To see this we employ the following 
set of $\gamma$ matrices
\begin{eqnarray}
  \gamma^j=\sigma^j\otimes\sigma^1, \qquad
  \Gamma^1=\sigma^3\otimes\sigma^1, \qquad
  \Gamma^2=1\otimes\sigma^2. \qquad
  (j=1,2)
\end{eqnarray}
The $\mathrm{spin}(2)$ generator and chiral 
matrix is given by 
\begin{eqnarray}
  \Sigma_3=\Sigma^{12}
  =\frac{i}{2}\sigma^3\otimes\sigma^3, \qquad
  \gamma_5=(-i)^2\gamma^1\gamma^2\Gamma^1\Gamma^2
  =1\otimes\sigma^3. 
\end{eqnarray}

Zero mode wave function can be chosen to be chiral.
We write it in chiral spinors $\psi_\pm$ as
\begin{eqnarray}
  \psi_\pm(x)&=&\left(
    \begin{matrix}
      u_\pm(x) \\ v_\pm(x)
    \end{matrix}\right)
\otimes\chi_\pm,
\end{eqnarray}
where $\chi_\pm$ are two component eigenspinors 
with $\sigma^3\chi_\pm=\pm\chi_\pm$.  
Each component of the chiral zero mode satisfies 
in polar coordinates 
\begin{eqnarray}
  \left\{-ie^{i\theta}\left(\partial_r+\frac{i}{r}\partial_\theta\right)
    \pm i\frac{1-k}{r}e^{i\theta}\right\}u_\pm
  -g|\phi_0|he^{\mp i\theta}v_\pm&=&0, \nonumber \\
  g|\phi_0|he^{\pm i\theta}u_\pm
  +\left\{-ie^{-i\theta}\left(\partial_r-\frac{i}{r}\partial_\theta\right)
      \pm i\frac{1-k}{r}e^{-i\theta}\right\}v_\pm&=&0,
\end{eqnarray}
where we have introduced Yukawa coupling 
constant $g$ by replacing $\phi$ with $g\phi$ in 
Eq. (\ref{eq:gsh}). The index (\ref{eq:indtd}) is not 
affected by this change as mentioned in Sect. \ref{sec:dga}. 

\begin{figure}[t]
  \centering
  \epsfig{file=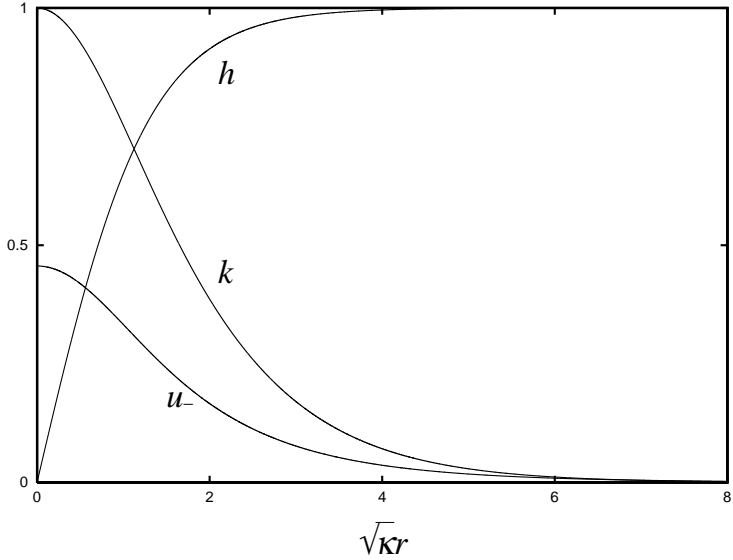,clip=,height=8cm,angle=0}  
  \caption{Profiles of  $h(r)$, $k(r)$ and $u_-(r)$ with $\lambda=\kappa$ 
and $g=e/2$.}
  \label{fig:vortex}
\end{figure}

For the negative chirality zero mode we can assume 
that $u_-$ and $v_-$ are independent of $\theta$. 
It is easy to check that a normalizable solution 
is given by 
\begin{eqnarray}
  u_-=-i \mathrm{sgn}(g)v_-=C_0\exp\left[-\int_0^r
    \left(|g\phi_0|h(r')+\frac{1-k(r')}{r'}\right)dr'\right],
\end{eqnarray}
where $C_0$ is a normalization constant. 
In Figure \ref{fig:vortex} we give a plot for the profile 
of $u_-(r)$ together with $h(r)$ and $k(r)$. The zero mode wave 
function is localized around the vortex core.

As for the positive chirality zero mode, we can separate 
the angle variable by assuming  
\begin{eqnarray}
  u_+(x)=f_m(r)e^{im\theta}, \qquad
  v_+(x)=ig_m(r)e^{i(m+2)\theta},
\end{eqnarray}
where $m$ is an integer. 
$f_m$ and $g_m$ satisfy 
\begin{eqnarray}
  &&\left(\frac{d}{dr}-\frac{m}{r}-\frac{1-k}{r}\right)f_m
  +g|\phi_0|hg_m=0, \nonumber \\
  &&\left(\frac{d}{dr}+\frac{m+2}{r}-\frac{1-k}{r}\right)g_m
  +g|\phi_0|hf_m=0.
\end{eqnarray}
These lead to the behaviors $f_m\sim r^m$ and $g_m\sim r^{-m-2}$ 
as $r\rightarrow0$. We see that only one of the two independent 
solution is regular at the origin. Such a regular solution, 
however, contains exponentially growing component 
$\sim e^{|\phi_0|r}$ as $r\rightarrow\infty$. 
We thus conclude that there is no positive chirality 
zero mode. 

\subsection{'t Hooft-Polyakov monopole}
\label{sec:tpmb}

Next we consider a Yang-Mills-Higgs system with $\mathrm{spin}(3)$ 
gauge symmetry in three spatial dimensions. 
The ansatz for the monopole of unit magnetic charge is given by  
\begin{eqnarray}
  \label{eq:mppA}
  \hat\phi_a(x)=h(r)\frac{x^a}{r}, \qquad
  A_{abi}(x)=-(1-k(r))\frac{\delta_{ia}x^b-\delta_{ib}x^a}{r^2}.
\end{eqnarray}
Eqs. (\ref{eq:sfeq}) are satisfied if $h(r)$ and $k(r)$ 
obey the following differential equations
\begin{eqnarray}
  \label{eq:hk}
  h''+\frac{2}{r}h'=\frac{2}{r^2}k^2h-\lambda(1-h^2)h, \qquad
  k''=\kappa h^2k-\frac{1}{r^2}(1-k^2)k. 
\end{eqnarray}
The boundary conditions for $h(r)$ and $k(r)$ take the 
same form as Eq. (\ref{eq:BChkv}) for the vortex. 
From Eqs. (\ref{eq:hk}) we see the asymptotic behavior 
$k \sim e^{-\sqrt{\kappa}r}$ and 
$1-h\sim e^{-\sqrt{2\lambda} r}$ for sufficiently large $r$. 
No analytic solution is not known for Eqs. (\ref{eq:hk}). 
See Ref. \cite{for} for a recent high precision numerical study. 

It is straightforward to evaluate $\mathrm{index}H$. In 
odd dimensions only chiral current Eq. (\ref{eq:Jithd}) 
contributes to the index. Note that the covariant derivatives 
$D_i\hat\phi$ decays exponentially as $r\rightarrow\infty$ and 
the field strength approaches
\begin{eqnarray}
  F_{abij}\rightarrow
  \frac{\delta_{ia}\delta_{jb}}{r^2}
  -\frac{\delta_{ia}x^jx^b
    -\delta_{ib}x^jx^a}{r^4}-(i\leftrightarrow j).
\end{eqnarray}
Keeping terms that survive at $r\rightarrow\infty$, we obtain 
\begin{eqnarray}
  J^i\approx-\frac{1}{8\pi}\mathrm{sgn}(g)\epsilon^{ijk}\epsilon^{abc}
  \hat\phi_aF_{jkbc}=-\mathrm{sgn}(g)\frac{x^i}{2\pi r^3},
\end{eqnarray}
where the overall factor $\mathrm{sgn}(g)$ comes from the introduction of 
the Yukawa coupling constant. In odd dimensions the index depends 
on the sign of $g$. This immediately gives $\mathrm{index}H=\mathrm{sgn}(g)$. As in the 
vortex case, it is also possible to obtain the same result 
by computing the topological invariant (\ref{eq:indthd}). 

\begin{figure}[t]
  \centering
  \epsfig{file=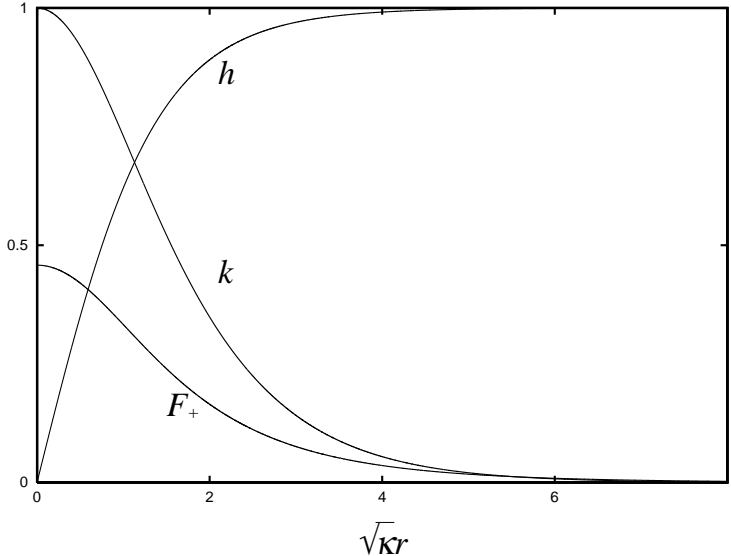,clip=,height=8cm,angle=0}  
  \caption{Profiles of $h(r)$, $k(r)$ and $F_+(r)$ for the monopole 
background with $\lambda=\kappa/2$ and $g=e/2$.}
  \label{fig:mp}
\end{figure}

The index obtaind above implies the existence of 
a zero mode of chirality $\mathrm{sgn}(g)$. For the monopole background 
Eq. (\ref{eq:mppA}) it is also possible to find the 
wave function for the zero mode. 
We employ the following representation of the $\gamma$ matrices
\begin{eqnarray}
  \gamma^i=\sigma^i\otimes1\otimes\sigma^1, \qquad
  \Gamma^a=1\otimes\sigma^a\otimes\sigma^2. \qquad(i,a=1,2,3) 
\end{eqnarray}
The $\mathrm{spin}(3)$ generators $\Sigma_a=\frac{1}{2}\epsilon_{abc}
\Sigma^{ab}$ and the chiral matrix $\gamma_7$ are given by
\begin{eqnarray}
  \Sigma_a=\frac{i}{2}1\otimes\sigma^a\otimes1, \qquad 
  \gamma_7=1\otimes1\otimes\sigma^3. 
\end{eqnarray}
Let us denote the zero mode wave function by  
chiral components as 
\begin{eqnarray}
  \psi&=&\left(
    \begin{matrix}
      \psi_+ \\ \psi_-
    \end{matrix}\right). 
\end{eqnarray}
Then $\psi_\pm$ must satisfy 
\begin{eqnarray}
  \label{eq:zmeq}
  \left(-i\sigma^j\otimes1\partial_j
      +\frac{1}{2}\sigma^j\otimes\sigma^aA_{aj}
      \pm ig1\otimes\sigma^a\phi_a\right)\psi_\pm&=&0. 
\end{eqnarray}
where $A_{ai}$ is defined by 
$A_{ai}=\frac{1}{2}\epsilon_{abc}A_{bci}$.
These can be cast into $2\times2$ matrix equations 
by noting $(A\otimes B\psi_\pm)_{\alpha\beta}=A_{\alpha\gamma}B_{\beta\delta}
  (\psi_\pm)_{\gamma\delta}
  =(A\psi_\pm B^T)_{\alpha\beta}$. 
With this notation Eq. (\ref{eq:zmeq}) for the 
monopole background Eq.(\ref{eq:mppA}) can be expressed 
as 
\begin{eqnarray}
  -i\sigma^j\partial_j\Psi_\pm-\epsilon_{jab}(1-k)\frac{x^b}{2r^2}\sigma^j
  \Psi_\pm\sigma^a\mp ig|\phi_0|h \frac{x^a}{r}\Psi_\pm \sigma^a&=&0. 
\end{eqnarray}
where $\Psi_\pm$ are defined by $\Psi_\pm=i\psi_\pm\sigma^2$. 
These have spherically symmetric solutions
\begin{eqnarray}
  (\Psi_\pm(x))_{\alpha\beta}&=&F_\pm(r)\delta_{\alpha\beta},
\end{eqnarray}
where $F_\pm$ satisfy
\begin{eqnarray}
  F_\pm'(r)&=&-\left(\pm g|\phi_0|h(r)+\frac{1-k(r)}{r}\right)F_\pm(r). 
\end{eqnarray}
For $g>0$ the negative chiral component $F_-$ must vanish, otherwise 
it grows exponentially as $r\rightarrow\infty$. We thus arrive at 
the normalizable positive chiral zero mode 
\begin{eqnarray}
  &&\psi=\left(
    \begin{matrix}
      iF_+(r)\sigma^2 \\ 0
    \end{matrix}\right), 
\end{eqnarray}
where $F_+$ is given by 
\begin{eqnarray}  
  F_+&=&C_0\exp\left[-\int_0^r\left(g|\phi_0|h(r')
      +\frac{1-k(r')}{r'}\right)dr'\right].
\end{eqnarray}
Again $C_0$ is a normalization constant. 
The zero mode is localized around the monopole and 
the wave function decays exponentially for $r\rightarrow\infty$.
In Figure \ref{fig:mp} we give a plot of $F_+(r)$ 
together with $h(r)$ and $k(r)$. 
The case of $g<0$ can be analyzed similarly. We obtain one 
normalizable zero mode with negative chirality. This is 
consistent with the index theorem. 

\section{Summary and Discussion}
\label{sec:sd}
\setcounter{equation}{0}

We have evaluated the index of BdGH of a 
gauged topological insulator or Yang-Mills-Higgs 
system in arbitrary dimensions by regarding 
the Higgs and Yang-Mills fields as external 
backgrounds, which can be set up independently. 
The index can be expressed as a 
surface integral of a gauge invariant chiral 
current plus topological index of the Yang-Mills 
fields. In odd dimensions the 
topological index vanishes identically and the 
gauge field dependent terms of the chiral current 
can be gathered into a total derivative at 
spatial infinities, giving no contribution to 
the index. In even dimensions the gauge field 
dependent terms of the chiral current can be 
converted into a total derivative plus the 
Chern-Simons form, which exactly cancels the 
topological index. We have thus shown that 
the index of the BdGH is determined solely by 
the asymptotic behavior of the Higgs fields 
whatever topological charge the Yang-Mills 
field carries. 

If the behavior of the Yang-Mills 
and Higgs fields are governed by some effective 
Hamiltonian, $D_i\phi_a$ and $F_{ij}$ must decay 
faster than $|x|^{-d/2}$ for $|x|\rightarrow\infty$ 
to ensure the finiteness of the Hamiltonian. 
In such systems nonvanishing topological 
invariant can be obtained only in spatial 
dimensions less than four. In two dimensions 
the index of the BdGH is saturated by the 
topological index. In three dimensions 
only the $\phi F$ term of the chiral current 
contributes to the index. This apparently 
contradicts to the general conclusion that 
the Yang-Mills field does not contribute to 
the index. It is , however, possible to have 
expressions for the index only in terms of 
$\phi$ by noting that the gauge fields are 
related with the Higgs fields by $D_j\hat\phi_a=0$ 
at the spatial infinities. 

We have considered the case of the BdGH containing 
$d$ order parameters from the beginning. It is 
possible to consider other systems with less 
or more order parameters. The evaluation of 
index of the corresponding BdGH is straightforward. 
It is rather obvious from our explicit calculations 
that one cannot obtain nontrivial index unless 
the number of the order parameters coincides 
with the spatial dimensions. Our result is 
consistent with the the topological classification 
by the Chern number computed from the Berry 
connection of the Bloch wave functions.

\section*{Acknowledgements}

This work is supported in part by the Grant-in-Aid for 
Scientific Research (No. 21540378) from the Japan Society 
for the Promotion of Science (JSPS) and by 
the ``Topological Quantum Phenomena'' 
Grant-in Aid for Scientific Research on Innovative 
Areas (No. 23103502) from the Ministry of Education, 
Culture, Sports, Science and Technology of Japan (MEXT).

\end{document}